\documentclass[onecolumn]{aa}
\usepackage{natbib}
\usepackage{graphicx}
\usepackage{txfonts}

\def\aj{AJ}                   % Astronomical Journal
\def\aap{A\&A}                % Astronomy and Astrophysics
\def\aaps{A\&AS}              % Astronomy and Astrophysics, Supplement
\def\apj{ApJ}                 % Astrophysical Journal
\def\apjl{ApJ}                % Astrophysical Journal, Letters
\def\pasp{PASP}               % Publ. of the Astron. Society of the Pacific
\def\procspie{Proc.SPIE}      % Proceedings of the SPIE
\newcommand{\etal}{et al.~}

\newcommand{\bet}{$\beta$~}
\newcommand{\micron}{$\mu$m}

\begin{document}
\title{HD 199143 and HD 358623: Two Recently Identified Members of the
  \bet Pictoris Moving Group} 

\author{D. Kaisler\inst{1}
\and 
B. Zuckerman\inst{1}
\and
I. Song\inst{1,2}
\and
B.A. Macintosh\inst{3}
\and
A. J. Weinberger\inst{4}
\and
E.E. Becklin\inst{1}
\and
Q. M. Konopacky\inst{1,3}
\and
J. Patience \inst{3,5}}

\offprints{D. Kaisler}

\institute{Department of Physics and Astronomy, UC Los Angeles
\and
Institute for Geophysics and Planetary Physics, UC Los 
Angeles
\and
Institute for Geophysics and Planetary Physics, Lawrence
Livermore National Labs
\and
Department of Terrestrial Magnetism, Carnegie
Institution of Washington
\and
Division of Physics, Mathematics, and Astronomy,
  California Institute of Technology
}

\titlerunning{HD 199143 and HD 358623}
\authorrunning{Kaisler et al.}

\date{Received xxx, 2003; accepted xxx, 2003}

\abstract{
HD 199143 and HD 358623 (BD-17$\degr$6128) are two sets of binary stars
which are physically associated and 48 pc from Earth. We present
{\bf heliocentric} radial velocities and high lithium abundances which establish
these stars as members of the $\sim$12 Myr-old \bet  Pictoris Moving Group. 
We also present mid-IR photometric measurements which show no firm
evidence for warm dust around all four stars.   

}

\maketitle

\section{Introduction} \label{intsec}

The discovery of an increasingly large number of young stars near Earth
is providing fertile ground for studies of Galactic star
formation, including the initial kinematics of the local stellar
population, the evolution of the stars themselves, and the formation of their
primordial disks. During the past two years, a number of authors have
presented data on the rapidly rotating F8V star HD 199143 and the
K7-M0 dwarf HD 358623. Based on their common proper motion,
\citet{van00} (hereafter vdA00) concluded that these stars are
associated and have an age of $\sim$ 20 Myr. To explain EUVE and ROSAT
\citep{vog99, vog00} detections, rapid rotation, variability in the
continuum and emission lines of HD 199143, vdA00 postulated a
protoplanetary disk or unseen stellar companion.  Abundant lithium (EW =
400 m\AA ; \citet{mat95}) in HD 358623 and chromospheric activity of both
stars also indicate young ages. \citet{van01} (vdA01) drew on the similar
proper motions of these two stars to classify them as a new young
cluster, which they named the Capricornus association.  

Adaptive optics (AO) imaging by Jayawardhana \& Brandeker (2001)
(hereafter, JB01) revealed companion candidates to HD 199143 and HD
358623 at separations of 1 and 2\arcsec~ respectively.  Astrometry by
\citet{neu02} firmly established HD 358623 B as a true companion of HD
358623 A. vdA01 measured N (11 \micron) and Q (19 \micron) band excesses
around both HD 199143 and HD 358623 implying circumstellar
disks. Near-IR photometry by JB01 resulted in an extremely red color
(J-K = 1.4) for HD 199143B which was interpreted as evidence of a
circumsecondary disk. However, \citet{cha02} later showed that this
photometry was highly dependent upon the deconvolution parameters and
suggested that this influenced the assignment of an infrared excess to
one of the stars. 

Despite these discoveries, the reported ages of these stars have
been inconsistent. Since age is the key factor in constraining models of
lithium depletion, dust disk dissipation, and planetary formation, it is
desirable to have an accurate estimate of the ages for these stars. In
this paper, we present near- and mid-infrared photometry as well as
optical spectroscopy to show that the space motions and spectral
features of HD 199143 and HD 358623 are consistent with those of the $\beta$
Pictoris Moving Group; from this, and from high lithium
abundances,  we infer young ages for the two systems. 

\section{Observations} \label{obssec}

\subsection{Near-IR Photometry with Adaptive Optics} \label{nirsec}

We observed HD 199143 and HD 358623 at four epochs (Table 1). The AO
system at the Shane 3-m telescope at Lick observatory was used with
IRCAL, a 256 $\times$ 256 Rockwell HgCdTe array with a plate scale of
0.0756 arcsec/pixel \citep{llo00}. Strehl ratios (at 2.1\micron) were in
the range 0.3--0.45.  Absolute photometry for HD 199143 and HD 358623
was obtained on 30 June 2002 by comparing open-loop images with the
standard star HD 201941 \citep{eli82}. Measurements were done with IRAF
using an aperture radius of 40 pixels (3\arcsec). Although the K'and
K$_s$ filters used for this study have slightly shorter bandpasses
than the \citet{eli82} K-band filter, \citet{per98} find that the 
color-correction between K and K$_s$ is $<$ 0.02 mag even for stars as red
as HD 199143 and HD 358623. Therefore, we have made no color correction
here, {\bf but have added in quadrature an additional 0.02 mag error to
  the quoted K magnitude (Table 2).}

%%%%%%%%%%%%%%%%%%%%%%%%%%%%%%%%%%%%%%%%%%%%%%%%%%%%%%%%%%%%%%%%%%%%%%%%%%%%%
%%%% Table 1

\begin{table}
\centering 
\caption[]{Near-Infrared Observing Log \label{obstbl}}
\vspace*{0.2cm}
\begin{tabular}{llll}
\hline
\hline
\\
Target   &  Epoch \& UT            &  Filter & Instrument \\
\\
\hline
\\
HD 199143&  2001-06-14    & Ks & IRCAL\\
         &  2001-06-14    & J  & `` \\

         &  2001-07-03    & K$^{\prime}$ & SCAM \\
         &  2001-09-29    & Ks & IRCAL\\
         &  2001-09-30    & J  & IRCAL\\ 
         &  2002-06-30    & Ks & IRCAL\\
         &  2002-06-30    & J  & ``\\
HD 358623&  2001-06-14    & Ks & IRCAL\\
         &  2001-06-14    & J  & ``\\
         &  2001-07-03    & K$^{\prime}$ & SCAM\\
         &  2002-06-30    & Ks & IRCAL\\
         &  2002-06-30    & J  & ``\\
\\
\hline
\hline
\end{tabular}
\end{table}

%%%%%%%%%%%%%%%%%%%%%%%%%%%%%%%%%%%%%%%%%%%%%%%%%%%%%%%%%%%%%%%%%%%%%%%%%%%%%

At the Keck II telescope, we used the AO system with NIRSPEC's
slit-viewing camera, SCAM \citep{mcl98}. The plate scale for SCAM's 256
$\times$ 256 HgCdTe array is 0\farcs$0167\pm$0\farcs001/pixel.
Strehl ratios for this run (K$^{\prime}$-band) were 0.10--0.46.

Photometric measurements of HD 199143 AB were performed using Starfinder
routine \citep{dio00} and are reported in Table 2. Our estimates
of the magnitudes of these stars agree well with those of other studies
except in the case of the JB01 J-band measurement of HD 199143
B. However, \citet{cha02} note that the JB01 value may be brought into
agreement with the others by the choice of a more robust myopic deconvolution
algorithm. 

On 14 June 2001, we measured the binary separations and position angles
as 1\farcs02 $\pm$ 0\farcs03 and -36\fdg0 $\pm$ 0\fdg5 ~for HD 199143 and
2\farcs12 $\pm$ 0\farcs02 and 137\fdg9 $\pm$ 0\fdg5 ~for HD 358623. Our
position angles differ slightly from those published by JB01, however 
due to the superior weather conditions under which our observations were
made \citep{cha02}, we believe that our measurements are more accurate.

%%%%%%%%%%%%%%%%%%%%%%%%%%%%%%%%%%%%%%%%%%%%%%%%%%%%%%%%%%%%%%%%%%%%%%%%%%%%%
%%%% Table  2

\begin{table} \label{photbl}
\centering
\caption[]{Near-IR Photometry}
\vspace*{0.2cm}
\begin{tabular}{llll} 
\hline
\hline
\\
Star & J & K & J-K \\
\\
\hline
\\
HD 199143 A & 6.24$\pm0.03$ & 5.89$\pm0.04$ & 0.35$\pm0.05$ \\
HD 199143 B & 8.92$\pm0.07$ & 8.01$\pm0.07$ & 0.91$\pm0.10$ \\
HD 358623 A & 7.92$\pm0.03$ & 7.12$\pm0.04$ & 0.80$\pm0.05$ \\
HD 358623 B & 9.67$\pm0.05$ & 8.77$\pm0.06$ & 0.90$\pm0.08$ \\
\\
\hline
\hline
\end{tabular}
\end{table}

%%%%%%%%%%%%%%%%%%%%%%%%%%%%%%%%%%%%%%%%%%%%%%%%%%%%%%%%%%%%%%%%%%%%%%%%%%%%%

\begin{figure} 
\centering
\includegraphics[width=0.50\columnwidth]{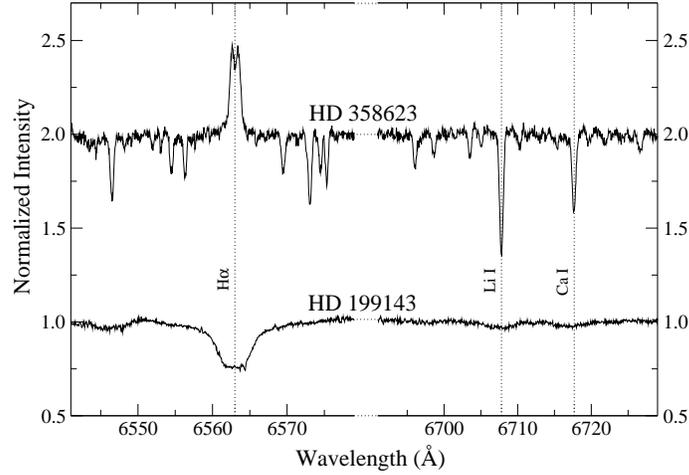}
\caption{Hamilton Echelle spectra of HD 199143 and HD 358623. \vspace{1.0cm}}
\end{figure}

\begin{figure}
\centering
\includegraphics[width=0.50\columnwidth]{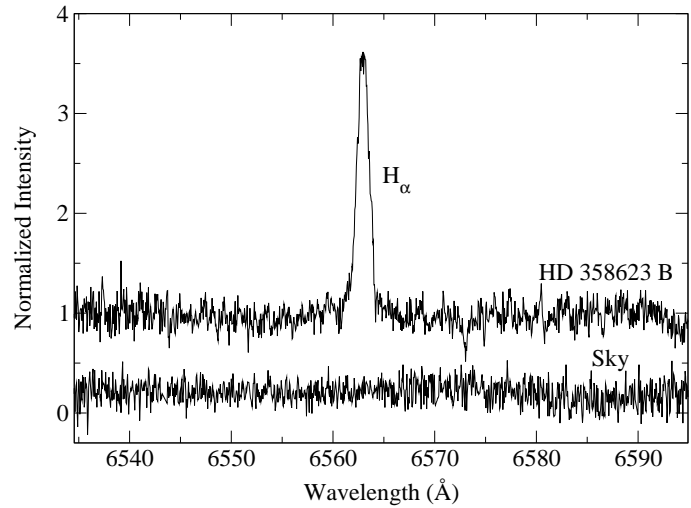}
\caption{Hamilton spectra of HD 358623 B. To estimate the       
contamination from the primary, we took an empty sky spectrum with the
same separation from the primary and opposite position angle. Equivalent   
width of the H$\alpha$ line is $\sim 4$ \AA.}  
\end{figure}

\subsection{Optical Spectroscopy} \label{specsec}

We obtained high-resolution ($\sim0.15$ \AA) spectra of HD 199143 and
HD 358623 (Fig. 1) with the Hamilton Echelle spectrometer \citep{vog87} at the
Shane Telescope. HD 358623 was observed on UT 18 June 2001 and both HD
stars were observed on 29 July 2001. Seeing was $\sim$ 1\arcsec
~on both nights.  Spectra were reduced using IRAF following a standard
procedure (flat-field correction, etc.). Equivalent widths of H$\alpha$
and Li6708 features were measured using the IRAF task, {\bf
splot}. Terrestrial atmospheric features were removed using an A-type
rapid rotator, HR 7235.

The heliocentric radial velocity of HD 358623 (Table 3) was measured by
cross-correlating its spectrum with radial velocity standards HR 911, HR
6349, HR 8232, and HR 8969. All standards gave consistent results. The
weighted mean of the measured radial velocities of HD 358623 is
-6.8$\pm$1.3 km/s. 

To verify that our H$\alpha$ spectrum of HD 358623 B was uncontaminated by
light from HD 358623, we placed the slit first on B and then on
a blank field at equal distance from and opposite position angle with
respect to A. Figure 2 confirms that there is no scattered light from
the primary star at this distance. The projected dimensions of the slit
used for this observation are 1.5 arcsec wide in the dispersion
direction and 2.5 arcsec long in the spatial direction.

%%%%%%%%%%%%%%%%%%%%%%%%%%%%%%%%%%%%%%%%%%%%%%%%%%%%%%%%%%%%%%%%%%%%%%%%%%%%%
%%%% Table 3

\begin{table} \label{rvtbl}
\centering
\caption[]{Results From Optical Spectroscopy}
\vspace*{0.2cm}
\begin{tabular}{lllll}
\hline
\hline
\\
Target & EW(H$\alpha$) & EW(Li) & vsin(i) &  v$_{r}$  \\
 & [{\AA}] & [{m\AA}] & [km/s] & [km/s]  \\
\\
\hline
\\
HD 199143 A  &  1.6  & 120: & 120$^a$  & -- \\
HD 358623 A  &  -0.78$^b$ & 400$\pm$20  & 13  & -6.8 $\pm$ 1.3 \\
HD 358623 B  &  $\sim$-4$^b$ & -- & -- & -- \\
\\
\hline
\hline
%\vspace{0.5cm}
\multicolumn{5}{l}{a : Uncertainty $\sim$ 10 - 20 \%.}\\
\multicolumn{5}{l}{b : Emission line.}\\
\end{tabular}
\end{table}

\subsection{Mid-infrared Photometry} \label{mirsec}

Mid-infrared observations of HD 199143 and HD 368623 were taken on 7 Dec
and 21 Dec 2001 at the W. M. Keck Observatory with the facility Long
Wavelength Spectrograph (LWS) \citep{jon93}. LWS's 128 $\times$ 128 
pixel array has a plate scale of 0.08 arcsec pixel$^{-1}$ and a
focal-plane field of view of 10\farcs2 square. The conditions were
photometric during both sets of observations. The image quality was
limited by seeing to FWHM of 0\farcs5 at 12 $\mu$m and 0\farcs6 at 18
$\mu$m on 7 Dec and 0\farcs8 at 8--12 $\mu$m and 0\farcs6 at 18 $\mu$m
on 21 Dec. 

The images were taken in a chop-nod mode with a chop amplitude of
10\arcsec, a nod of 10\arcsec\ in the same direction, and a chopper
frequency of 5 Hz. This scheme produces four images, two of which contain
the object and two of which contain background only. The four images were
double differenced for optimal background subtraction of the sky and
telescope thermal emission. Filter characteristics are tied to the IRAS
calibration.

The photometric standard star HR 8728 (Fomalhaut) was observed near in
time and close in airmass to the observations of the HD stars. From its
color-corrected IRAS flux densities, we calculate Fomalhaut's
mid-infrared magnitude as 0.88 mag from 8--18 $\mu$m with an uncertainty
of 10\%. On 21 Dec, observations of other standards at smaller airmasses
are consistent with this magnitude and uncertainty for HR 8728 given
typical Mauna Kea airmass corrections.

For HD 199143, the flux ratio of the binary was found by PSF fitting at 12
$\mu$m to be 5.6 $\pm$ 0.1 and the total flux density from both components
was measured in a 4\farcs8 synthetic aperture. The 18 $\mu$m image had
insufficient signal-to-noise ratio for the same PSF fitting procedure to
work, so the flux densities were measured using small, diameter 1\arcsec,
aperture photometry around each star. The total flux density in a very
large aperture agreed with the sum in the small ones. For HD 358253, the
photometry of each component was found in a synthetic aperture of
diameter 1\arcsec\ with residual background measured in an annulus
3\farcs5 from the source and subtracted. Statistical uncertainties
calculated from the background were added in quadrature. The total
uncertanties are reported in Table 4. Column two of this table
represents the total integration time.

%%%%%%%%%%%%%%%%%%%%%%%%%%%%%%%%%%%%%%%%%%%%%%%%%%%%%%%%%%%%%%%%%%%%%%%%%
%%%% Table 4

\begin{table} 
\centering
\caption[]{Mid-infrared Observations and Results \label{midirobs}}
\vspace*{0.2cm}
\begin{tabular}{lcccc}
\hline
\hline
\\
Filter     & t$_{int}$ & Airmass & primary & secondary \\
\micron    & (s)       &        &(mJy)   & (mJy) \\
\\
\hline
\\
\multicolumn{5}{c}{HD 199143}\\
11.7    &216   &1.53  &144 $\pm$ 16  &32 $\pm$ 4 \\
17.65   &240   &1.55  &66  $\pm$ 25  &41 $\pm$ 14 \\
\multicolumn{5}{c}{HD 358623}\\
8.9     &144   &1.90  &62  $\pm$ 9   &34 $\pm$ 7   \\
11.7    &240   &2.2   &35  $\pm$ 5   &15 $\pm$ 4 \\
17.65   &360   &2.1   &$<$20         &$<$20\\
\\
\hline
\hline
\end{tabular}
\end{table}

%%%%%%%%%%%%%%%%%%%%%%%%%%%%%%%%%%%%%%%%%%%%%%%%%%%%%%%%%%%%%%%%%%%%%%%%%

\subsection{ROSAT and Near-infrared Imaging}

\citet{vog00} report a bright (0.28 cts/s) X-ray source about one arc
minute north of HD 199143 which itself is listed in the ROSAT All-Sky
Survey.  We examined ROSAT images of the field around HD
199143. With 30\arcsec~ pixelation, we see no indication at all of an
X-ray source one minute north of the star.  We also investigated
the area north of the star with UCLA's twin-channel Gemini camera
\citep{mcl94} at the Shane Telescope on 5 August 2001. A 60~s J-band
exposure with limiting magnitude (3$\sigma$) of 19.1 mag reveals no
source 1\arcmin~ to the north of HD 199143. There is also no
indication on the corresponding ROSAT image of an X-ray source at the
listed position. 

\section{Results and Discussion}

We use our photometry to place these stars in a color-magnitude diagram
along isochrones by \citet{bar02} (Figure 3). All four stars lie on or
to the right of the 20 Myr isochrone.

To investigate the reliability of Baraffe \etal isochrones at low
masses, we selected 29 K and M-type dwarfs with less than 10\% errors in
parallax from the \citet{gle69} and \citet{gle91} catalogues. When added
to the color-magnitude diagram (Fig. 3), these field stars, which we
expect to be of main sequence age, parallel the Baraffe 100 Myr
isochrone. This is consistent with a pre-main-sequence phase for the HD
199143 and HD 358623 pairs. 

\begin{figure*}
\centering
\includegraphics[width=\columnwidth]{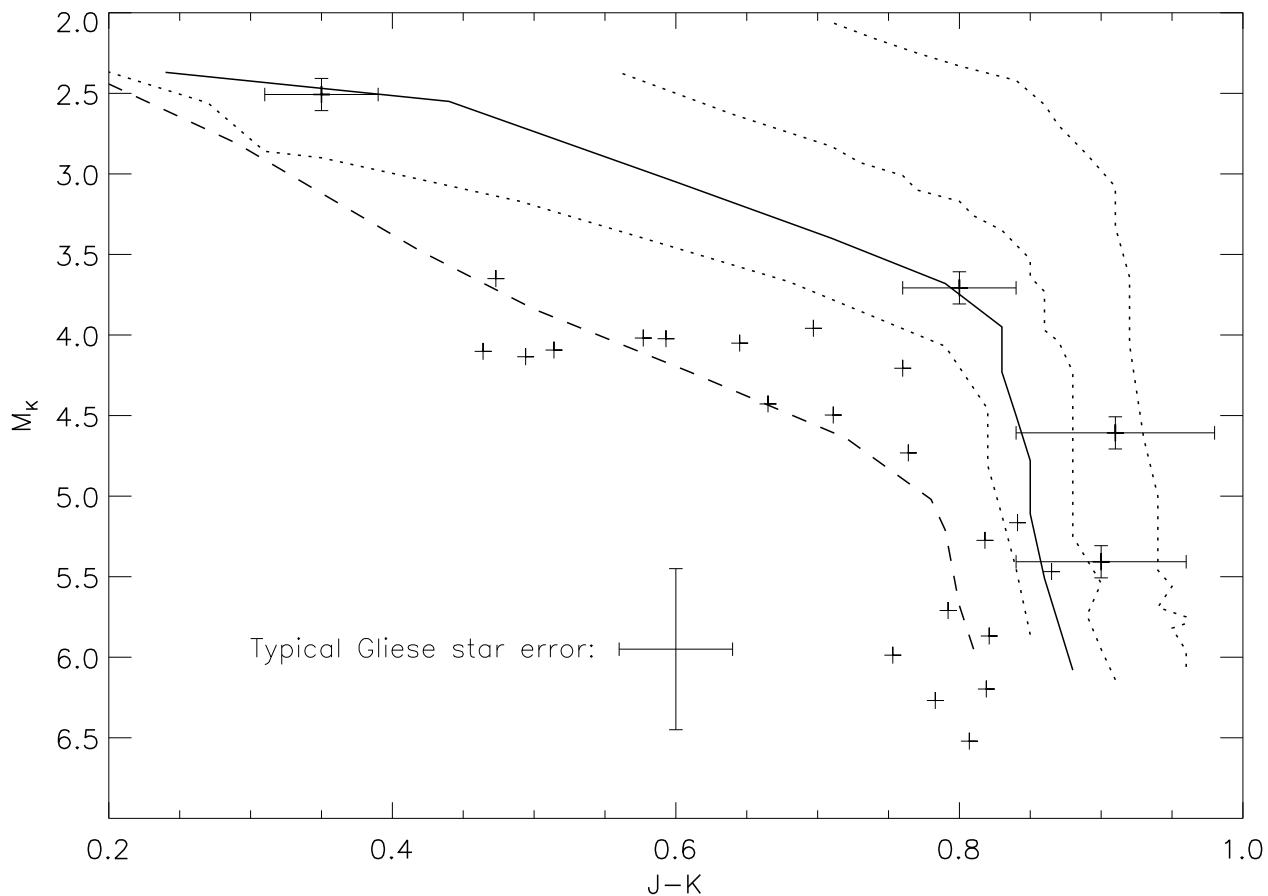}
\caption{Color-magnitude diagram showing (top to bottom) HD 199143 A, HD
358623 A, HD 199143 B, and HD 358623 B. \citet{bar02} ~isochrones,
from left to right, are 100 Myr (dashed line) 32 Myr, 20 Myr (solid line),
10 Myr, and 3 Myr. Positions of main-sequence Gliese stars are
indicated by the '+' symbols. Colors are from the 2MASS catalogue.} 
\end{figure*}

Further evidence for the youth of HD 358623 comes from the 400 m\AA
~equivalent width of its Li $\lambda$6708 absorption line. Figure 1
of \citet{son02} -- a plot of (B-V) color vs. Li $\lambda$6708 EW -- shows that
all stars with equivalent widths $\sim$400 m\AA ~belong to very young
groups such as the TW Hydrae association, or the open cluster NGC 2264.  

The ages of HD 199143 and HD 358623 are consistent with the
12$^{+8}_{-4}$ age derived by \citet{zuc01} for members of the $\beta$
Pictoris Moving Group (BPMG) and confirmed by \citet{ort02}, who
obtained a kinematic age of 11.5 Myr by tracing the orbits of all
proposed BPMG members backwards in time through a 3-D Galactic
potential. Furthermore, the Galactic space motion of HD 358623,  
relative to the Sun: (-9.8$\pm$0.6, -15.5$\pm$0.3, -11.1$\pm$0.5),
derived using Hipparcos distances and proper motions, is consistent with
the mean motions (-10.5, -15.8, -9.4) \citep{ort02} and velocity dispersions
($\pm \sim$2km/sec in each direction) of the published BPMG stars
\citep{zuc01}.     
 
Figures 4 (HD 199143) and 5 (HD 358623) show synthetic spectra fit to
the near-infrared photometry. In all four stars, the mid-infrared flux
densities are consistent with those expected from the photospheres. These 
results contradict the measurements of vdA00 and vdA01 by showing no
substantial infrared excess. The small excess that appears around HD
199143B at 18$\mu$m is only significant at the 2$\sigma$ level.  

\begin{figure}
\vspace{1.5cm}
\centering
\includegraphics[width=0.50\columnwidth]{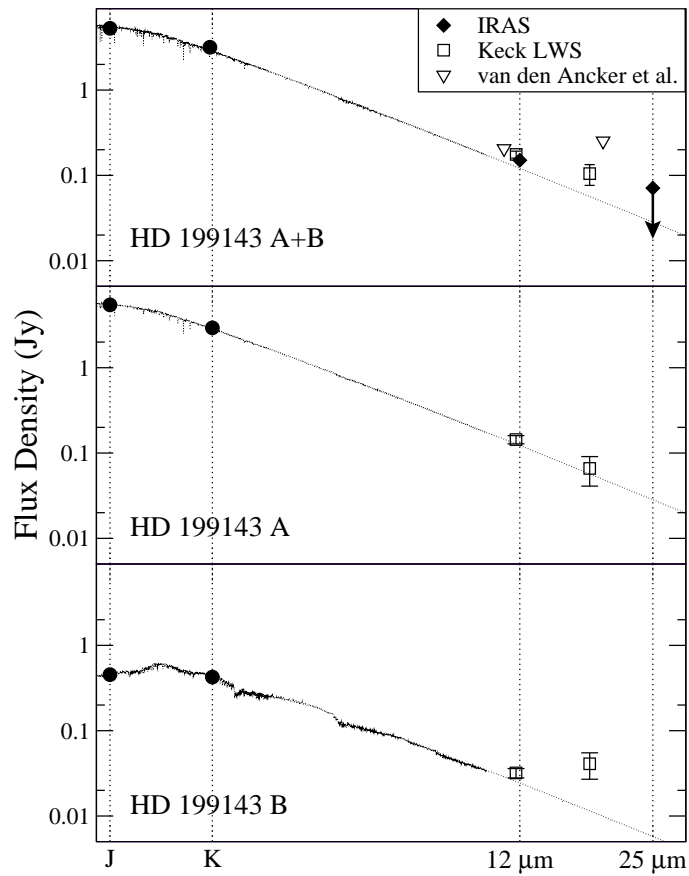}
\vspace{1.cm}
\caption{Synthetic SEDs \citep{hau99} for HD 199143 A and B with
overplotted mid-IR data. Effective temperatures for the models are
$T_\mathrm{eff}$(A) = 6400 K and $T_\mathrm{eff}$(B) = 3700 K, with 
Z = 0.02 and log(g) = 4.5 for both models. Filled circles represent
the photometric measurements of Section 2.1.}
\label{sedfig1}
\end{figure}

\begin{figure}
\vspace{1.5cm}
\centering
\includegraphics[width=0.50\columnwidth]{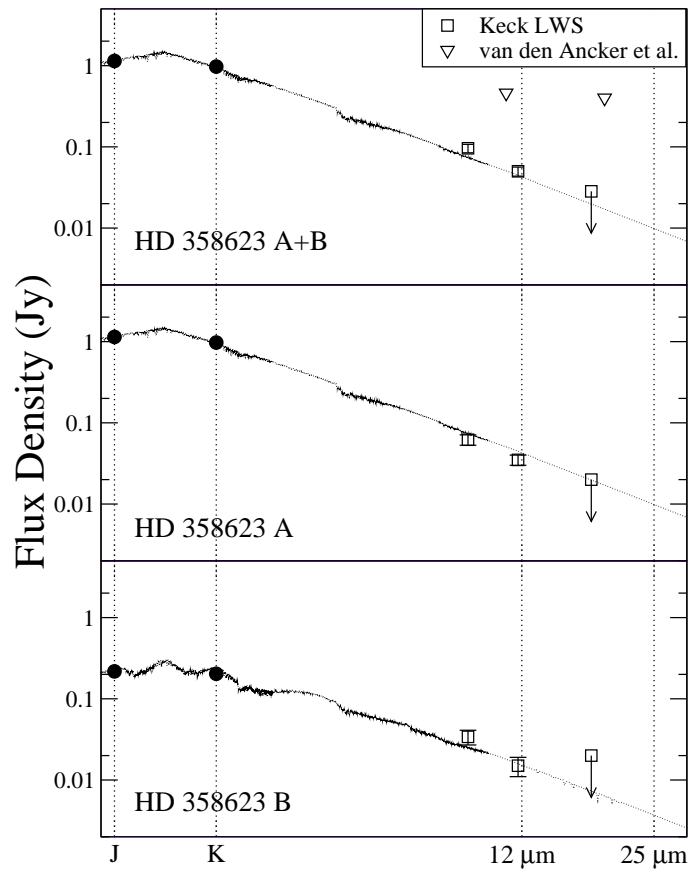}
\vspace{1.cm}
\caption{Synthetic SEDs \citep{hau99} for HD 358623 A and B with
overplotted mid-IR data. $T_\mathrm{eff}$(A) = 4400 K and
$T_\mathrm{eff}$(B) = 3300~K, with Z = 0.02 and log(g) = 4.5 for both
models. The plotted upper limits at 17.65\micron ~are 1-sigma
values. Filled circles represent the photometric measurements of section
2.1. }
\label{sedfig2}
\end{figure}

\section{Conclusions} \label{concsec}

Data presented in this report show that HD 199143 AB and HD 358623 AB
are members of the \bet Pictoris Moving Group, rather than a separate
Capricornus group. The high   lithium abundance of HD 358623 suggests an
age of less than roughly 10 Myr for this star. Near-IR adaptive optics
photometry and models by \citet{bar02} also suggests a young age (less
than 20 Myr). Our near- and mid-IR photometric results show no firm
evidence for the presence of warm dust around either star. Investigation
of a listed ROSAT All-Sky Survey source one arcminute north of HD 199143
revealed no near-IR counterpart to a limiting magnitude of J = 19.1 mag.

\acknowledgements{We would like to thank Randy Campbell, Jay Farihi, and
Patrick Lowrance for their useful input to this study, as well as the
referees, Drs. A. Brandeker and R. Liseau. This research was supported in part
by a NASA grant to UCLA and by NASA's Astrobiology Institute. Some of
the data presented herein were obtained at the W.M. Keck Observatory,
which is operated as a scientific partnership among the California
Institute of Technology, the University of California, and the National
Aeronautics and Space administration. The Observatory was made possible
by the generous financial support of the W.M. Keck foundation.}

\end{document}